\documentclass[sigconf]{acmart}

\usepackage{hyperref}
\usepackage[utf8]{inputenc}
\usepackage{balance}
\AtBeginDocument{%
  }

\setcopyright{none}
\copyrightyear{2018}
\acmYear{2018}
\acmDOI{XXXXXXX.XXXXXXX}

\acmConference[Conference acronym 'XX]{Make sure to enter the correct
  conference title from your rights confirmation email}{June 03--05,
  2018}{Woodstock, NY}
\acmISBN{978-1-4503-XXXX-X/18/06}

\begin{document}

\ccsdesc[500]{Human-centered computing~Virtual reality}
\ccsdesc[500]{Human-centered computing~Mixed / augmented reality}
\title[Exploratory AI-driven Visualisation Techniques in XR]{An Exploratory Study on AI-driven Visualisation Techniques on Decision Making in Extended Reality}

\author{Ze Dong}
\authornotemark[1]
\email{ze.dong@pg.canterbury.ac.nz}
\affiliation{%
  \institution{School of Product Design, University of Canterbury}
  \city{Christchurch}
  \state{Canterbury}
  \country{NZ}
}

\author{Binyang Han}
\authornotemark[2]
\email{binyang.han@pg.canterbury.ac.nz}
\affiliation{%
  \institution{School of Product Design, University of Canterbury}
  \city{Christchurch}
  \state{Canterbury}
  \country{NZ}
}

\author{Jingjing Zhang}
\authornotemark[3]
\email{jingjing.zhang@pg.canterbury.ac.nz}
\affiliation{%
  \institution{School of Product Design, University of Canterbury}
  \city{Christchurch}
  \state{Canterbury}
  \country{NZ}
}

\author{Ruoyu Wen}
\authornotemark[4]
\email{ruoyu.wen@pg.canterbury.ac.nz}
\affiliation{%
  \institution{School of Product Design, University of Canterbury}
  \city{Christchurch}
  \state{Canterbury}
  \country{NZ}
}

\author{Barrett Ens}
\authornotemark[5]
\email{barrett.ens@ubc.ca}
\affiliation{%
  \institution{Department of Computer Science, The University of British Columbia}
  \city{Vancouver}
  \state{BC}
  \country{Canada}
}

\author{Adrian Clark}
\authornotemark[6]
\email{adrian.clark@canterbury.ac.nz}
\affiliation{%
  \institution{School of Product Design, University of Canterbury}
  \city{Christchurch}
  \state{Canterbury}
  \country{NZ}
}

\author{Tham Piumsomboon}
\authornotemark[7]
\email{tham.piumsomboon@canterbury.ac.nz}
\affiliation{%
  \institution{School of Product Design, University of Canterbury}
  \city{Christchurch}
  \state{Canterbury}
  \country{NZ}
}

\renewcommand{\shortauthors}{Ze Dong}
\renewcommand{\shortauthors}{Binyang Han}
\renewcommand{\shortauthors}{Jingjing Zhang}
\renewcommand{\shortauthors}{Ruoyu Wen}
\renewcommand{\shortauthors}{Barrett Ens}
\renewcommand{\shortauthors}{Adrian Clark}
\renewcommand{\shortauthors}{Tham Piumsomboon}


\begin{abstract}
  The integration of extended reality (XR) with artificial intelligence (AI) introduces a new paradigm for user interaction, enabling AI to perceive user intent, stimulate the senses, and influence decision-making. We explored the impact of four AI-driven visualisation techniques---`Inform,' `Nudge,' `Recommend,' and `Instruct'---on user decision-making in XR using the Meta Quest Pro. To test these techniques, we used a pre-recorded 360$^{\circ}$ video of a supermarket, overlaying each technique through a virtual interface. We aimed to investigate how these different visualisation techniques with different levels of user autonomy impact preferences and decision-making. An exploratory study with semi-structured interviews provided feedback and design recommendations. Our findings emphasise the importance of maintaining user autonomy, enhancing AI transparency to build trust, and considering context in visualisation design.
  
\end{abstract}



\keywords{Extended Reality, Artificial Intelligence, Decision Making, Visualisation Techniques.}


\begin{teaserfigure}
  \centering
  \includegraphics[width=\textwidth]{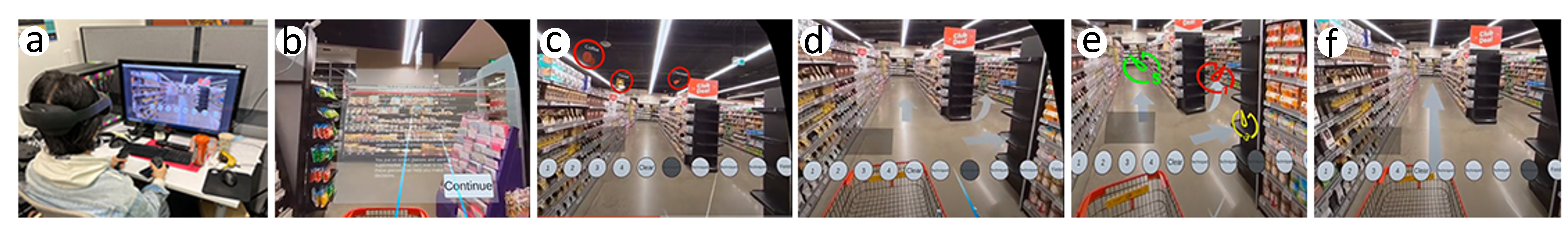} 
  \caption{The exploratory study setup and screenshots of the experiences: a) experimental setup, b) introduction to Event 1 (E1), c) `Inform' in E1, d) `Nudge' in E1, e) `Recommend' in E1, and f) `Instruct' in E1.}
  \Description{Teaser Figure}
  \label{fig:teaser}
\end{teaserfigure}

\maketitle

\section{Introduction}

As artificial intelligence (AI) continues to advance, previous research has explored its role in automating, optimising, and enhancing the generation and interpretation of visual data, aiding users in comprehending complex information more effectively \cite{wu2021ai4vis}. AI's data analysis and pattern recognition capabilities have been shown to significantly improve decision-making accuracy and efficiency \cite{chandana2023impact}. However, there is still considerable potential to explore how AI can be more effectively harnessed to understand user intentions and create visualisations that better align with user needs, thereby enhancing decision-making processes in real-time. Through the classification of AI autonomy by Parasuraman et al.\cite{parasuraman_model_2000} and O'Neill et al.\cite{oneill_humanautonomy_2022}, we see the potential for leveraging different levels of AI autonomy to drive visualisations that are more tailored to user intentions while preserving varying degrees of user autonomy.

Augmented reality (AR) glasses, as key devices within the extended reality (XR) platform---exemplified by models like XReal Air 2 \footnote{\url{https://www.xreal.com/air2}} and Vuzix Z100 \footnote{\url{https://www.vuzix.com/products/z100-smart-glasses}}---offer environmental adaptability, real-time information overlay, and multimodal content presentation. These features highlight the potential for enhanced visualisation techniques when AI and XR are combined. This integration could establish a new paradigm of human-computer interaction, significantly influencing decision-making processes. In the initial stages of our study, we simulate the functionalities of AR glasses by using 360-degree video in Virtual Reality (VR). By integrating virtual elements into these videos, we emulate the real-world experience of AR glasses within a controlled VR environment. This simulation allows us to test our design ideas and explore effective design paradigm for AI-driven visualisation techniques in XR. By doing so, we aim to identify optimal ways to enhance user decision-making processes. We believe as AR glasses become more prevalent and integrated into daily life, the opportunities for using AI-driven visualisation in XR to assist decision-making are expected to be more pervasive.

Our motivation is to address the challenge of aligning AI-driven visualisation techniques in XR with user needs. Our preliminary research focuses on two aspects: First, we held an interdisciplinary workshop to identify research gaps, design considerations, and how to display visualisations based on different levels of AI autonomy. Based on the workshop's findings, we developed a user interface (UI) design and tested it in an exploratory study. We simulated a UI for AR glasses using a 360$^{\circ}$ video of a supermarket with a virtual interface overlay to investigate user preferences and feedback across different events in the same setting. This study evaluated how AI-driven visualisation techniques, customised for specific scenarios, can meet user decision-making needs, laying the foundation for future research in designing and implementing these techniques in XR.

\section{Related work}

\subsection{AI-driven Visualisation Techniques for Decision Support in XR}

Previous research has examined varying levels of AI involvement in decision-making \cite{parasuraman_model_2000, oneill_humanautonomy_2022, araujo_ai_2020, ma_beyond_2024}. Neill et al. \cite{oneill_humanautonomy_2022}, building on Parasuraman et al. \cite{parasuraman_model_2000}, categorised agent autonomy into low, medium, and high levels. Despite the advancements in AI-driven XR systems, there is still limited understanding of how different levels of AI-driven visualisation techniques influence user behaviour and decision-making. Addressing this gap is the central focus of our ongoing research, which aims to delineate the impact of these techniques, guided by prior studies \cite{parasuraman_model_2000, oneill_humanautonomy_2022, piumsomboon_ex-cit_2022}.

\begin{figure*}[htbp]
\includegraphics[width=\linewidth]{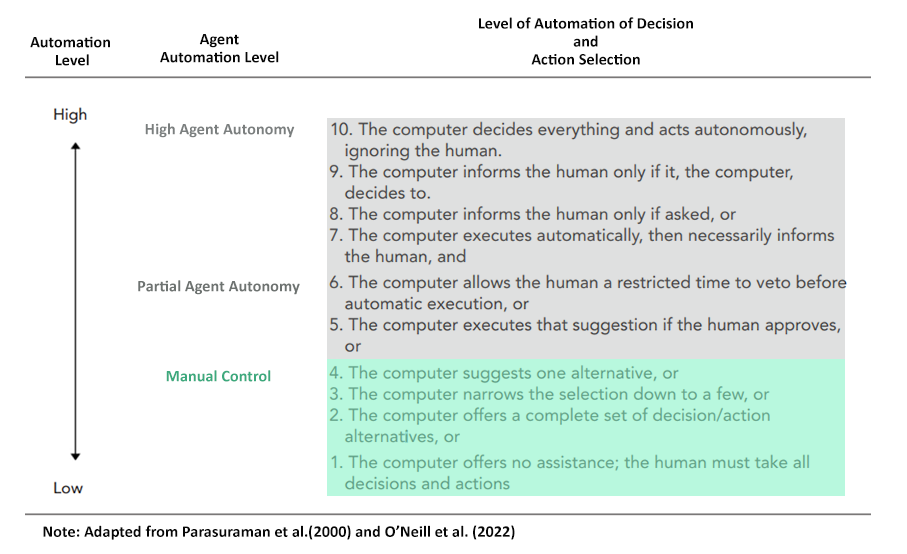}
\caption{The agent autonomay level adapted by Parasuraman et al. \cite{parasuraman_model_2000} and O'Neill et al. \cite{oneill_humanautonomy_2022}}
\Description{The AI autonomy Level}
\label{fig:1}
\end{figure*}

As data complexity and volume increase, effective data management becomes challenging, often leading to missed opportunities, wasted resources, and financial losses. Appropriate information visualisation techniques can enhance decision-making by improving data comprehension and communication \cite{keim2008visual}.

The integration of AI with XR significantly improves how data is processed and presented, aiding decision-making \cite{reiners_combination_2021}. Martins et al. \cite{martins2022augmented} explored the integration of AR and AI within Decision Support Systems (DSS), emphasising the role of situated visualisation in enhancing decision-making. Their study demonstrated how AR can provide real-time, context-sensitive visual data overlays, offering intuitive and immersive insights that improve decision-making effectiveness and efficiency.

Expanding on these developments, Kim et al. \cite{kim_interaction_2023} demonstrated the utility of AR visual cues in autonomous driving, enhancing safety and user cognition. Xu et al. \cite{xu_xair_2023} investigated the use of explainable AI (XAI) in AR-based intelligent service interactions, proposing a design framework for integrating these technologies. Sadeghi et al. \cite{sadeghi_virtual_2021} highlighted the potential of XR and AI in visualising complex structures, while Liu et al. \cite{liu_interactive_2022} and Mahmud et al. \cite{mahmud_visual_2023} explored enhancements in usability and accessibility of virtual environments through visual cues.

In sports training, AI-driven visualisation in XR has shown potential for improving decision-making and technical skills. Tsai et al. \cite{tsai2021feasibility} and Chen et al. \cite{chen2023shootpro} demonstrated how these techniques can enhance decision-making speed and provide immediate feedback in basketball, offering scalable and cost-effective training solutions.

Recent hardware advancements, such as the AR glasses by Viture and XReal \cite{viture_viture_2024, xreal_xreal_2024}, have made AI-compatible XR interfaces more affordable, enriching user experiences \cite{yoon_opportunities_2021, abrash_creating_2021}. PANDALens, a Proactive AI Narrative Documentation Assistant, exemplifies this by integrating with AR glasses to enhance daily activities through intelligent assistance \cite{demPandalens, pandalens}. It uses multimodal contextual information to generate coherent narratives with minimal user effort, improving both writing quality and travel enjoyment.

AI-driven visualisations within XR have the potential to revolutionise decision-making across various fields by offering dynamic and immersive ways to manage and interpret complex information. This research aims to explore the effects of different levels of AI-driven visualisation techniques in XR on user decision-making in daily shopping scenarios. An exploratory study will observe user behaviours and preferences, investigate the reasons behind these behaviours, and reassess user evaluations of these techniques. This work contributes to the knowledge on AI-driven visualisation technologies for decision support in XR, providing insights for optimising these technologies to enhance decision-making.

\subsection{Human-AI Interactions and Their Qualitative Analysis}

Human-AI interaction, a specialised field within HCI \cite{liu2023human}, involves AI systems that integrate hardware and software to simulate human cognitive behaviours such as creativity, learning, and speech. These systems operate autonomously in complex environments, solving problems that surpass human capabilities due to data complexity. This interaction views AI systems as intelligent entities with independent behaviours, extending the traditional scope of HCI \cite{wienrich2021extended}.

The integration of XR and AI marks a new interaction paradigm, overlaying virtual interfaces onto the user's surroundings and enabling voice-enabled interaction. This fusion enhances communication with AI, helping to better interpret user intent, stimulate the senses, and influence decision-making and behaviour \cite{reiners_combination_2021,hirzle_when_2023,qayyum_secure_2024}. As this technology advances, the need for explainable, accountable, and intelligible AI systems becomes critical. Abdul et al. \cite{abdul2018trends} emphasise designing AI systems that are both technically effective and understandable to build user trust and comprehension.

Further exploring human-AI interaction, Bansal et al. \cite{bansal2021does} discuss AI-generated explanations in decision-making within human-AI teams. Their findings indicate that while explanations increase trust and acceptance, they do not necessarily improve team performance, especially when AI errors occur. Complementing this, Roy et al. \cite{roy2019automation} highlight that high controllability in AI systems boosts user satisfaction, even with less accurate automation, underscoring the importance of user control.

Lu et al. \cite{lu_exploring_2023} investigate how user traits, such as personality and trust propensity, along with AI performance and transparency, affect human-AI interactions in head-mounted displays (HMD) during AI-assisted spatial tasks. They suggest that future AI assistance on HMDs should consider individual user characteristics and customise system design accordingly.

Qualitative analysis is vital for understanding user needs and behaviours in human-AI interaction. It offers deeper insights into how users interact with technology, revealing nuanced responses beyond simple binary answers \cite{blandford2016qualitative}. For instance, Zhu et al. \cite{ZHUhaiQA2023100041} used qualitative methods to uncover how a lack of transparency and incomprehensible information in AI systems, such as Robo-advisors, can lead to distrust and hinder adoption.

Similarly, Siemon’s research employs qualitative interviews to identify key roles for AI in enhancing team dynamics, showing AI's potential as a collaborative team member \cite{mixedmethod}. Farič et al. \cite{farivc2024early} underscore the importance of qualitative research by examining the integration of an AI-based diagnostic system in radiology. Their study reveals context-specific insights that aid in adopting AI technologies in sensitive environments like healthcare.

In conclusion, incorporating qualitative analysis into human-AI interaction research is crucial for designing systems that align with human needs and societal values. It deepens our understanding of how AI systems are perceived and used in real-world settings, supporting the development of user-centred and socially aware AI solutions \cite{papakyriakopoulos2021qualitative}. This paper contributes by systematically collecting and coding interview results, revealing behaviour patterns and preferences among participants, and providing insights into AI-driven visualisation techniques in XR. These findings are valuable for future design development and optimising user experiences.

\section{Interdisciplinary Workshop}
The workshop focused on exploring AI-driven visualisation in XR, with unexpected daily events chosen as the theme to set the context. The main goal of the workshop was to collaboratively explore and design AI-driven visualisation techniques in XR that support user decision-making during unexpected daily events. Approved by the Human Ethics Committee at the University of Canterbury (Ref: HREC 2024/51/LR-PS
), the workshop brought together experts in psychology, interaction design, AI, and XR to review current technologies, identify triggers, brainstorm solutions, develop design concepts, and evaluate their effectiveness.

\subsection{Participants}
Eight participants (2 females, 6 males, average age 32, SD=5.8) from the university took part, including three associate professors in HCI, XR, and AI, a postgraduate psychology student, two interaction design experts, a game designer, and a computer science research assistant.

\subsection{Procedure}

We used Edward de Bono’s Six Thinking Hats method \cite{de2017six} to structure the workshop, guiding strategic thinking on AI-driven visualisation in XR. The procedure integrated diverse perspectives, generating design ideas and an implementation agenda for an exploratory study focused on enhancing visualisation techniques. The method was applied as follows:

\begin{itemize}
    \item {\textbf{Blue Hat}: Management and Control. Introduction, icebreaker, and overview of AI-driven visualisation in XR.}
    \item {\textbf{White Hat}: Information and Facts. Discussion on the characteristics of unexpected events, research sharing, and identifying triggers in daily routines.}
    \item {\textbf{Yellow Hat}: Optimism and Positive Thinking. Brainstorming XR applications for managing unexpected events and related stress.}
    \item {\textbf{Black Hat}: Critical Judgement. Evaluation of potential problems, challenges, and limitations in AI-driven visualisation approaches.}
    \item {\textbf{Green Hat}: Creativity and New Ideas. Exploring novel AI-driven visualisation approaches in XR for managing unexpected events.}
    \item {\textbf{Red Hat}: Emotions and Feelings. Sharing initial reactions and conducting peer evaluations.}
    \item {\textbf{Blue Hat}: Management and Control. Summary, future research directions, and post-workshop survey.}
\end{itemize}

Our brainstorming focused on AI-driven visualisation techniques for AR glasses to aid decision-making under various AI capabilities.

\subsection{Results}
The workshop primarily focused on AI-driven visualisation techniques, particularly within the context of everyday AR glasses use. While several topics were discussed, we concentrated on two emerging themes: the level of AI capability, especially its autonomy in decision-making, and scenarios related to everyday activities, including potential unexpected events and associated stressors.

\subsubsection{Level of AI Autonomy}


The workshop findings aligned closely with the frameworks proposed by Parasuraman et al. \cite{parasuraman_model_2000} and O'Neill et al. \cite{oneill_humanautonomy_2022} on AI autonomy, which outline various levels of automation and human-AI interaction. These frameworks encapsulated key aspects of our discussions and provided a foundational basis for our design approach, particularly in emphasising human autonomy in decision-making processes. Building upon these established frameworks, we developed a tailored set of four levels of AI capability specifically for XR interfaces aimed at supporting decision-making while preserving user autonomy. Our proposed levels are:

\begin{itemize}
    \item {\textbf{Level 1---Inform:} Provides facts and information to aid decision-making \textit{(AI role-no assistance; human makes all decisions)}.}
    \item {\textbf{Level 2---Nudge:} Gently guides users towards beneficial choices \textit{(AI role-offers decision/action alternatives)}.}
    \item {\textbf{Level 3---Recommend:} Suggests and justifies options, encouraging informed decisions \textit{(AI role-narrows selection to a few, ranking them)}.}
    \item {\textbf{Level 4---Instruct:} Directs users with step-by-step instructions \textit{(AI role-suggests one alternative)}.}
\end{itemize}

We use the terms \textit{Nudge'} and \textit{Recommend'}, inspired by previous XR visualisation research on manipulating user perceptions \cite{piumsomboon_ex-cit_2022}. \textit{`Instruct'} was designed for emergencies where clear guidance is critical. Our visualisation techniques align with these levels, ranging from simple information displays to direct instructions, offering assistance in decision-making. Each level considers three factors: empowering the user by providing information and guiding decision-making, adapting to the situation's urgency or complexity, and respecting user autonomy, except in emergencies where the highest level of assistance is necessary.

\subsubsection{Stressors and Scenarios}
We identified common stressors from discussions and previous research \cite{wu_international_2015,phillips-wren_decision_2020}, focusing on three primary factors: time, finance, and health. These factors were central in our ideation phase and led to the creation of six pre-designed events (see table \ref{tab:six events}), shaping the study environment for AI-driven visualisation techniques.

Following Eichhorn et al. \cite{eichhorn_shopping_2023}, we chose supermarket shopping as a key scenario to elicit realistic responses. To simulate real decision-making, we used a pre-recorded 360$^{\circ}$ video from an actual supermarket, incorporating the six events influenced by the identified stressors. We hypothesise these will significantly impact decision-making in our test scenarios.


\begin{table*}[h]
\centering
\caption{The six events were identified after narrowing down the ideas generated during the interdisciplinary workshop.}
\begin{tabular}{|l|l|ll|}
\hline
\multicolumn{1}{|l|}{\textbf{Events}} & \textbf{Descriptions}                    & \multicolumn{2}{l|}{\textbf{Stressors}}\\ \hline
1. Unanticipated                 & Rearrangement of essential aisles             & \multicolumn{2}{l|}{Time}              \\ \hline
2. Contingency                   & Products out of stock                         & \multicolumn{2}{l|}{Finance \& Health} \\ \hline
3. Opportunity                   & Flash sale and clearance                      & \multicolumn{2}{l|}{Finance \& Health }\\ \hline              
4. Evaluation                    & Product comparisons on different platforms    & \multicolumn{2}{l|}{Finance \& Time}   \\ \hline
5. Disruption                    & Fire Drill                                    & \multicolumn{2}{l|}{Time}              \\ \hline
6. Interruption                  & Unexpected call                               & \multicolumn{2}{l|}{Time}              \\ \hline
\end{tabular}
\label{tab:six events}
\end{table*}


\section{Exploratory Study}
The aim of this study is to test the practicality of our design ideas in real-world supermarket settings. This stage will identify gaps between expected and actual user interactions through participant feedback and observations. Post-study semi-structured interviews will offer deeper insights into participants’ thoughts and preferences. These insights will help refine our data collection, management, and analysis methodologies. The findings will guide our UI design and implementation, preparing for detailed analysis in the next phase.

\subsection{Design of Visualisation Techniques with 360° Video}

Based on the workshop findings, our design of each visualisation technique considers three specific factors:

\begin{enumerate}
    \item \textbf{User Autonomy:} Allowing users to control the information they receive and how they interact with it, offering varying levels of assistance from basic information to detailed guidance.
    \item \textbf{Contextualisation:} Adapting the UI to different scenarios to ensure the information is relevant and useful in various contexts.
    \item \textbf{Progression:} Gradually increasing the complexity and detail of the information provided, enabling users to compare and understand the evolution across different visualisation levels.
\end{enumerate}

\begin{figure*}[htbp]
\includegraphics[width=\linewidth]{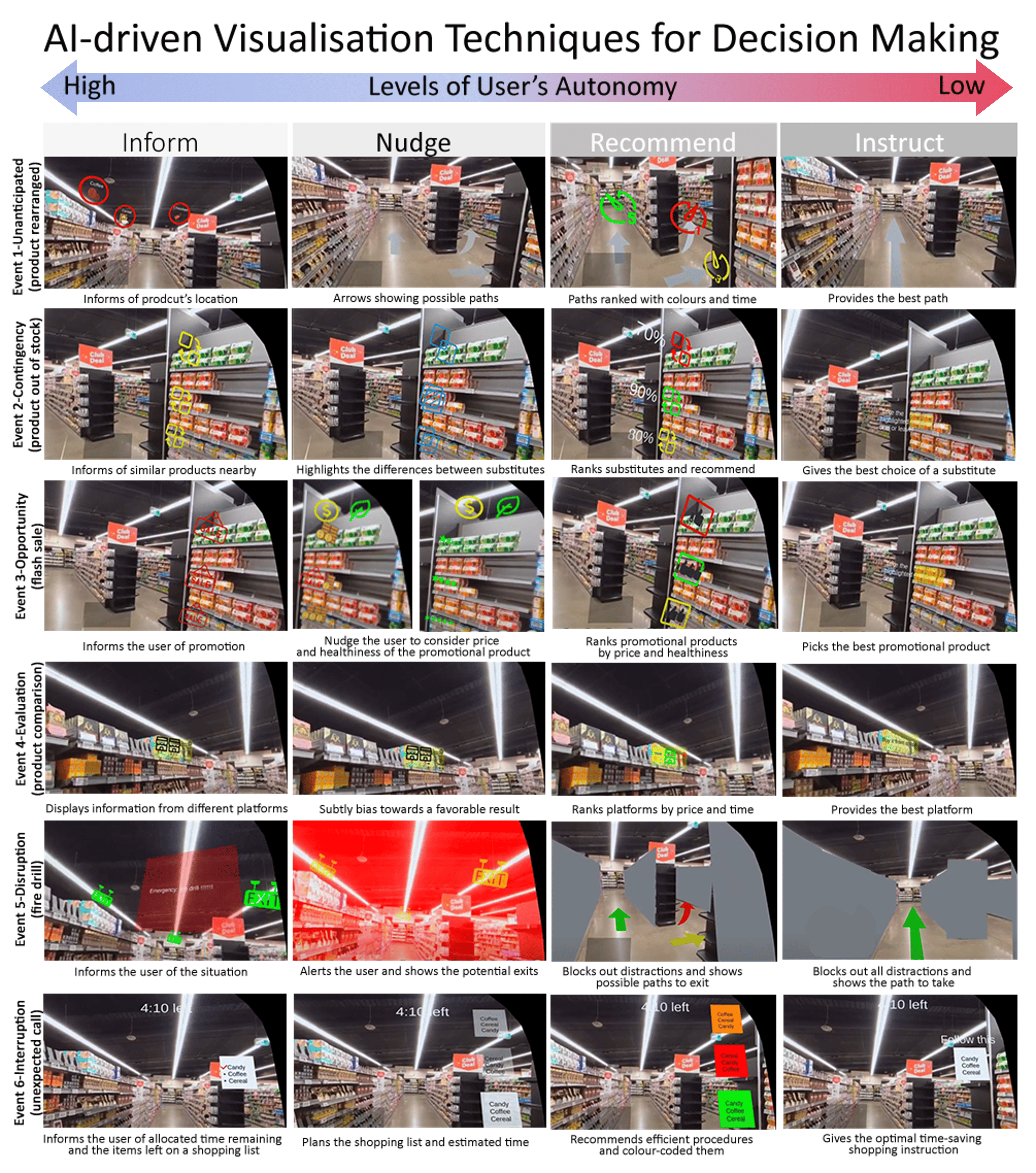}
\caption{All four visualisation techniques, Inform, Nudge, Recommend, and Instruct, across six events.}
\Description{System visualisation tech implementation}
\label{fig:2}
\end{figure*}

These factors ensure the designs align with the purpose of each technique and clarify how AI-driven visualisation influences decision-making. Figure \ref{fig:2} illustrates the UI elements for the four visualisation techniques overlaid onto a 360$^{\circ}$ video recorded in a supermarket, simulating the AR glasses experience across six events detailed in Table \ref{tab:six events}. 

\textbf{Event 1---Unanticipated:} When a supermarket rearranges its aisles, AI visualisation techniques help customers find products. Four techniques are used based on time: \textit{Inform}, displaying icons for product locations; \textit{Nudge}, guiding with flashing arrows and time estimates; \textit{Recommend}, showing ranked paths with colour-coded icons; and \textit{Instruct}, selecting the optimal path and directing the user.

\textbf{Event 2---Contingency:} If a product is out of stock, AI responds with visualisation techniques based on finance and health: \textit{Inform}, showing icons of similar products; \textit{Nudge}, comparing substitutes with overlapping icons; \textit{Recommend}, ranking alternatives with percentage values and colours; and \textit{Instruct}, calculating and highlighting the best alternative.

\textbf{Event 3---Opportunity:} When a promotion is discovered, AI's response includes: \textit{Inform}, notifying the user; \textit{Nudge}, evaluating products based on discount and healthiness; \textit{Recommend}, ranking items with colours and icons; and \textit{Instruct}, showcasing the best promotional choices.

\textbf{Event 4---Evaluation:} Addressing concerns about higher prices, AI assists by: \textit{Inform}, displaying prices and delivery times from different platforms; \textit{Nudge}, comparing these factors with symbols; \textit{Recommend}, ranking platforms based on price and delivery time; and \textit{Instruct}, highlighting the optimal purchasing platform.

\textbf{Event 5---Disruption:} During a fire drill, AI's visualisation focuses on safety: \textit{Inform}, indicating exits; \textit{Nudge}, tinting the view red and flashing exit icons; \textit{Recommend}, directing attention to escape paths with colour-coded arrows; and \textit{Instruct}, displaying the quickest exit route.

\textbf{Event 6---Interruption:} Handling an unexpected call requiring the user to meet soon, AI aids by: \textit{Inform}, updating on remaining time and shopping list; \textit{Nudge}, planning item collection order and updating time estimates; \textit{Recommend}, showing time-efficient shopping procedures; and \textit{Instruct}, optimising the shopping list order and maintaining time awareness.

\subsection{Participants}
We recruited eight postgraduate students from the university, evenly split by gender (four males, four females) with an average age of 28 years (SD=3.8). Half had XR experience, and all had AI experience, such as using ChatGPT \footnote{\url{https://chatgpt.com}}. Four participants self-assessed as having a basic understanding of AI, two had knowledge of AI-related terms, and two were involved in AI research. All were pre-informed about the study, participated voluntarily, and provided consent.

\subsection{Setup}
Our system was developed in Unity (2022.3.5f1) with AI-generated voices by ElevenLabs \footnote{\url{https://elevenlabs.io/}}. We used a 360$^{\circ}$ supermarket video, recorded with an Insta360 One R camera at $3840\times1920$ resolution, viewed through Meta Quest Pro \footnote{\url{https://www.meta.com/nz/quest/quest-pro/}}. The setup included a high-performance PC (Intel i7 8700, 3.2 GHz, 32 GB RAM, Nvidia GeForce RTX 3080). The study took place in a secure, quiet room to eliminate disturbances.

\subsection{Procedure}
After providing consent, participants shared their demographic information. They were then seated in the experimental space, where they learned to use the Meta Quest Pro and our system interface through a pre-recorded instructional video. Participants viewed various pre-recorded supermarket events, with the four visualisation techniques shown at specific points. The order of these techniques followed a Latin square design to prevent order effects. After experiencing all four techniques for each event, participants ranked them by preference and provided verbal explanations. This process was repeated for all six events. Following the events, participants participated in a semi-structured interview for more detailed feedback.

\begin{figure*}[ht]
\begin{center}
\includegraphics[width=\textwidth]{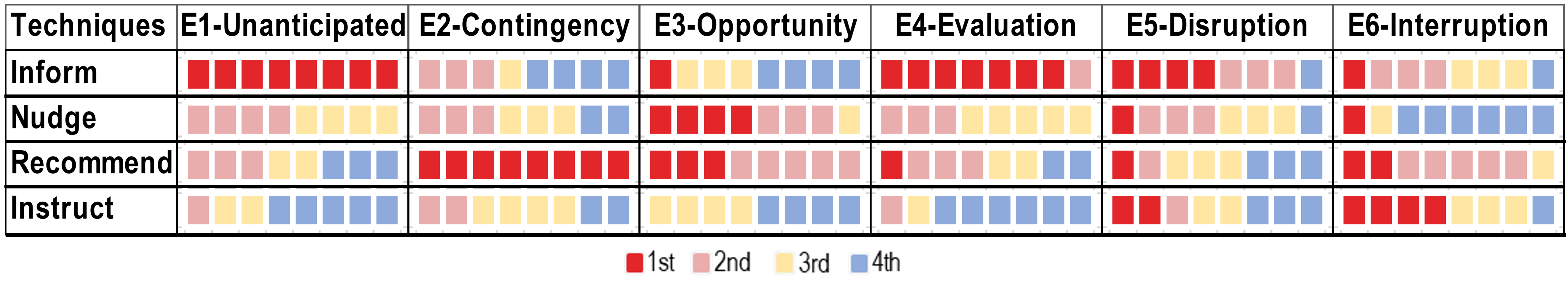}
\end{center}
\caption{The rankings in terms of user preferences for each visualisation technique for the six events.}
\Description{Preference}
\label{fig:3}
\end{figure*}

\section{Exploratory Study Results}
To understand user perceptions of visualisation techniques and their impact on various events and to refine our UI design, we conducted semi-structured interviews with 8 participants. These interviews explored their experiences and gathered feedback on their interactions with the simulated system. By coding the recordings, we identified key feedback, preferences, and patterns in their visualisation and interaction. This feedback provides valuable insights for the next stage of development. The report presents participants' preferences for each visualisation technique across events and categorises their feedback into three themes: system evaluation, information visualisation, and AI concerns.

\subsection{User Preferences}
The mean for the preferences of visualisation techniques are represented as \textit{Inform}, $\bar{x}_{if}, \textit{Nudge}, \bar{x}_{nu}, \textit{Recommend}, \bar{x}_{rc}, \textit{Instruct}, \bar{x}_{is}$, with lower numbers indicating higher participant preference, as shown in Figure \ref{fig:3}.

\textbf{Event 1 (Unanticipated):} The most preferred visualisation technique was in the following order: \textit{Inform} ({$\bar{x}_{if}$}=1, SD=0), \textit{Nudge} ({$\bar{x}_{nu}$}=2.5, SD=.54), \textit{Recommend} ({$\bar{x}_{rc}$}=3, SD=.93), and \textit{Instruct} ({$\bar{x}_{is}$}=3.5, SD=.76). 

\textbf{Event 2 (Contingency):} The ranking from most preferable visualisation tehchnique was \textit{Recommend} ({$\bar{x}_{rc}$}=1, SD=0), \textit{Nudge} ({$\bar{x}_{nu}$}=2.88, SD=.84), \textit{Instruct} ({$\bar{x}_{is}$}=3, SD=.76), and \textit{Inform} ({$\bar{x}_{if}$}=3.13, SD=0.99).

\textbf{Event 3 (Opportunity):}  The most preferred visualisation technique was \textit{Recommend}({$\bar{x}_{rc}$}=1.63, SD=.52), \textit{Nudge} ({$\bar{x}_{nu}$}=1.63, SD=.74), \textit{Inform} ({$\bar{x}_{if}$}=3.3, SD=1.04), and \textit{Instruct} ({$\bar{x}_{is}$}=3.5, SD=.54). 

\textbf{Event 4 (Evaluation):} The ranking from most preferable was \textit{Inform} ({$\bar{x}_{if}$}=1.125, SD=.354), \textit{Nudge} ({$\bar{x}_{nu}$}=2.63, SD=.52), \textit{Recommend} ({$\bar{x}_{rc}$}=2.63, SD=1.06), and \textit{Instruct} ({$\bar{x}_{is}$}=3.63, SD=.74). 

\textbf{Event 5 (Disruption):} The ranking was \textit{Inform} ({$\bar{x}_{if}$}=1.75, SD=1.035), \textit{Nudge} ({$\bar{x}_{nu}$}=2.5, SD=0.93), \textit{Instruct} ({$\bar{x}_{is}$}=2.75, SD=1.28), and \textit{Recommend} ({$\bar{x}_{rc}$}=3, SD=1.069).

\textbf{Event 6 (Interruption):} The rank was \textit{Recommend} ({$\bar{x}_{rc}$}=1.88, SD=.64), \textit{Instruct} ({$\bar{x}_{is}$}=2.13, SD=1.25), \textit{Inform} ({$\bar{x}_{if}$}=2.5, SD=0.93), and \textit{Nudge} ({$\bar{x}_{nu}$}=3.5, SD=1.07).

\subsection{User Feedback}

\subsubsection{Overall Design and Implementation}

Participants generally provided positive feedback on the system's design and implementation but highlighted the need for improved user adaptability and accessibility. They praised the UI concepts and found the application of AI-driven visualisation techniques in daily scenarios interesting. P1 and P7 noted that if AR glasses become as common as mobile phones, they could significantly change daily life.

P7, with a design background but no XR experience, emphasised the importance of user adaptability and adoption:

\begin{quote}
\textit{``It was very interesting to experience VR for the first time. However, I need more time to become familiar with the equipment and system. Your system should include a clear usage guide to prevent me from needing to ask you questions constantly.''}
\end{quote}

Participants appreciated the UI's forward-thinking design but raised concerns about the \textit{Instruct} technique potentially reducing shopping enjoyment and increasing stress in financial and health-related events. P6 remarked:

\begin{quote}
\textit{``Instruct technique may be useful at certain events, but most of the time, it reduces the interactivity with the system.''}
\end{quote}

Half the participants (P3, P5, P7, and P8) questioned the intended audience, noting that the UI design might not suit all users, especially elderly individuals and those with colour blindness. They suggested considering these factors to ensure accessibility and avoid added stress.

\subsubsection{Visualisation Techniques}

Participants suggested optimising the system's visualisation techniques for better clarity, control, and context awareness.. P4 and P7 noted that when the event's context was clear, the icons were easily understood. P4 added:

\begin{quote}
\textit{``When I know the background is a supermarket, I can easily understand the meaning of the corresponding icons under different visualisation techniques with context. I think these icons are appropriate in context, but the style is rigid. Your design may need to be improved, such as the location, colour, and text of the UI with context. Additionally, there is a lot of room for improvement in how the UI appears... Can dynamic UI be added? These factors should be considered in combination with visualisation techniques.''}
\end{quote}

Participants also provided specific feedback on each technique. P2, P3, and P4 wanted more control over the \textit{`Inform'} technique, seeking greater transparency in AI recommendations. For \textit{`Nudge'}, P7 and others felt it was too controlling, despite its clever prompting. The \textit{`Recommend'} technique was appreciated by P7 and P8 for aiding decision-making without being overwhelming. The \textit{`Instruct'} technique was deemed effective for quick decisions in emergencies (P1, P7).

P1 and P8 suggested adding more information during Event 5---Disruption, with P8 proposing a countdown timer to reduce stress. P1 added:

\begin{quote}
\textit{``I would be very inclined to use the system's `Recommend' and `Instruct' techniques if AI can inform me of the cause and severity of an emergency situation, and the pros and cons of an escape route in real-time and within a short period, it will provide help for find the most suitable escape route by avoiding crowded areas when everyone is running for their lives.''}
\end{quote}

P6 requested more detailed information during Event 4---Evaluation, particularly for financial decisions:

\begin{quote}
\textit{You should adjust the information detail of the corresponding visualisation techniques according to the event context. In event 4, I hope that the `Recommend' technique can provide more product information, such as the unit price. I don’t like the `Instruct' to tell me directly. I need more detailed information about prices and others.}
\end{quote}

Conversely, P5 preferred a concise, intuitive display, avoiding overstimulating colours and dynamic effects for comfort:

\begin{quote}
\textit{``I feel uncomfortable with the `Nudge' technique potentially aiding decision-making through flashing. The flashing induces a sense of tension. In an emergency situation, I need a more concise and intuitive visualisation technique to display information.''}
\end{quote}

P2 highlighted the need for simplified displays during Event 2---Contingency:

\begin{quote}
\textit{``The Contingency event display is more interesting. The use of icons allows me to obtain information more quickly. I dislike text and anything that requires reading.''}
\end{quote}

\subsubsection{Information Presentation}

We analysed participants' feedback on information display across different visualisation techniques. Participants emphasised the balance between detailed information and simplicity to enhance decision-making.. P3 suggested AI could assist by gathering product information from various sources, aiding in price comparisons:

\begin{quote}
\textit{``I am a budget person. If I can know the prices of other stores during my shopping, I can better manage my living costs, much like the Inform technique provided. Also, I would like to know the cost price of the product. The Inform technique should provide more information to better help me compare similar products and make informed decisions while shopping.''}
\end{quote}

Event 5---Disruption, a significant time stressor, was extensively discussed. Participants wanted clear escape routes and time prompts during emergencies. P2 noted:

\begin{quote}
\textit{``This scene impressed me the most. Under the 'Nudge' technique, my view turned completely red, which increased the sense of emergency. Although I knew the scene was fake, when immersed in it, I felt overwhelmed and couldn't distinguish the meaning and indicative nature of different UIs under this visualisation technique... The system's UI display should be clearer and more authoritative in emergency situations, similar to the 'Instruct' technique.''}
\end{quote}

Participants also suggested that more interactive design elements, like animations and conversation, could enhance engagement. P5, P6, and P8 recommended integrating these elements to reduce conflict and discomfort with UI overlays in the 360$^{\circ}$ video. Most participants (P1 to P5, P8) preferred informed assistance for better contextual understanding. The responsiveness of the interface to real-time conditions was highlighted as crucial. P5, P6, and P8 emphasised the need for varied visual cues, such as different animations, sizes, and transparency, to cater to diverse preferences and situations.

\subsubsection{AI Trust and Autonomy}

Participants' feedback revealed that trust in AI is complex and must be gradually built through actual usage, with concerns about over-reliance and maintaining user autonomy. P5 remarked:

\begin{quote}
\textit{``I don’t have many opportunities to use XR devices. I think the `Inform' technique is very effective, and AI does not intervene too much. Although the system provides a large amount of information sorted by AI, I have doubts about the ranking process. I am uncertain about how AI selects relevant information and I am concerned whether the ranking is influenced by manufacturers' advertising fees.''}
\end{quote}

P1 supported AI-driven recommendations but stressed the importance of retaining user autonomy for independent decisions. P4 expressed concerns about potential dependency on AI, noting:

\begin{quote}
\textit{``AI-driven Nudge' and Recommend' techniques are convenient and can facilitate better decision-making. But I am concerned that long-term use and reliance on the accuracy of AI-provided information may become a dependency problem. Then leading to a preference of users for `Instruct' in decision-making.''}
\end{quote}

Several participants (P4, P7, and P8) were concerned about AI's impact on autonomy and its potential to influence decisions. P3 highlighted the need for the AI system to manage multiple simultaneous events and provide appropriate assistance during complex situations.

\section{Discussion}
This section summarises our findings based on observations and user feedback, discussing their implications.

\subsection{Event-related Factors}
\textbf{Event 1---Unanticipated:} Participants strongly preferred the \textit{Inform} technique's straightforward approach, especially under time constraints, over more suggestive strategies like \textit{Nudge}, \textit{Recommend}, and \textit{Instruct}.

\textbf{Event 2---Contingency:} The \textit{Recommend} technique stood out, particularly under \textit{finance} and \textit{health} factors, indicating that users favour clear comparisons.

\textbf{Event 3---Opportunity:} Participants preferred \textit{Nudge} and \textit{Recommend} for unplanned events requiring decision support, suggesting that more detailed comparison and ranking are favoured in contexts involving \textit{finance} and \textit{health}.

\textbf{Event 4---Evaluation:} There was a strong preference for \textit{Inform} over low-autonomy techniques like \textit{Nudge} and \textit{Instruct}, emphasising the value of direct information under \textit{finance} and \textit{time} factors.

\textbf{Event 5---Disruption:} The lack of significant differences among techniques suggests they have similar impacts in disruption events, highlighting their potential interchangeability based on event characteristics.

\textbf{Event 6---Interruption:} Similar to Event 5, no significant differences were found among techniques, indicating their possible redundancy in interruption scenarios.

Overall, Events 1 and 4 show that straightforward alternatives like \textit{Inform} are sufficient for familiar tasks, while Events 2 and 3 demonstrate a preference for \textit{Recommend} and \textit{Nudge} in unfamiliar situations requiring additional decision support. This aligns with prior research \cite{xu_xair_2023} and suggests that AI explanations can enhance perceived autonomy and trust \cite{ali_explainable_2023}. The preference for each technique is context-dependent, varying with external factors.

\subsection{Design Implications}

\subsubsection{User Adaptability and Training}
Designing the UI should prioritise user adaptability to ensure a consistent experience, similar to adaptive UI improvements \cite{park2018adam,todi2021adapting}. The integration of XR and AI can enhance HCI by leveraging real-time data analysis. The system could dynamically monitor user focus, cognitive state, and environment, updating UI elements in AR glasses. For instance, during shopping, AI could deliver personalised information and recommendations. Intuitive visual cues and personalised tutorials can help users quickly adapt and improve usability.

\subsubsection{Visualisation Techniques and Decision-Making}
Colour coding \cite{singh2006impact,merenda2016effects} and other visualisation techniques \cite{ping2020effects,romano2023more} can enhance decision-making efficiency. Participants preferred comprehensive information, like prices and discounts, particularly in Event 3---Opportunity, for better comparison and decision-making. Future XR systems could allow users to compare AI-generated decision suggestions based on historical data and outcomes, aiding decision-making with the latest analytical results.

Beyond shopping, data sharing between emergency responders and patients via wireless communication can optimise emergency aid strategies. AI could detect real-time health data and display critical information on responders' AR glasses.

\subsubsection{User Autonomy and Trust in AI}



Participants emphasised the importance of maintaining autonomy despite AI's personalised recommendations \cite{calvo2020supporting,laitinen2021ai,miraz2021adaptive,evangelista2022auit}. Trust in AI systems is closely linked to their transparency; users are more likely to trust and accept AI recommendations when they understand how decisions are made \cite{yang2022user}. Transparent AI processes enable users to grasp the rationale behind recommendations, which is essential for building long-term trust. To enhance transparency, incorporating explainable AI (XAI) techniques can provide users with insights into the AI's decision-making processes \cite{xu_xair_2023}. By offering explanations through inspection mechanisms \cite{zicari2021z}, users can observe AI operations in real time via AR glasses, facilitating smooth transitions between manual and automatic processing. This real-time observation aligns with participants' desire for greater control and understanding of the AI system. Moreover, educational resources on AI and XR technologies can further enhance trust by demystifying complex AI functionalities and reducing apprehension towards automated assistance. By increasing users' knowledge and familiarity with the technology, they can make more informed decisions about when and how to rely on AI support.

Participants also highlighted that multimodal interfaces play a significant role in enhancing user engagement and autonomy. Utilising natural inputs like voice, gestures, and gaze control allows for more intuitive and seamless interactions with AI systems. These multimodal interfaces enable users to communicate with the system in ways that are comfortable and familiar, making it easier to override or modify AI recommendations when necessary.

\subsubsection{Diversity and AI-Assisted Collaboration}
Iterative optimisation of user experience design requires feedback from diverse participants \cite{adams2008qualititative}. Advanced visualisation techniques should offer effective information filtering and sorting in complex scenarios \cite{ellis2007taxonomy,lyi2020comparative}. 

Future XR platforms could simulate various study scenarios and remotely assemble diverse groups. AI could collect, analyse, and filter dynamic data in real time, improving user experience. In remote conferences, AI could organise relevant materials based on discussion topics and roles, facilitating collaboration. AI could also support asynchronous collaboration in XR through agents \cite{zhang2024virtual}.

\subsection{Limitations and Future Work}
As an exploratory study, our small sample size limits the generalisability of the results. However, we successfully identified potential research questions and hypotheses for future studies. In future iterations, we plan to scale up the study by involving a larger and more diverse participant pool, which will strengthen the validity of our findings and provide a more comprehensive understanding of user interactions with AI-driven visualisation techniques.  We also want to implement a working system for AI-driven visualisation techniques, moving beyond the current UI overlays. Based on participant feedback, we will refine the simplistic UI designs. This study only examined a supermarket setting, and many other use cases, such as healthcare,education and industrial environments remain to be explored. We hope this paper encourages further research into different scenarios.

\balance

\section{Conclusion}

This research involved an interdisciplinary workshop and an exploratory study to refine design ideas and establish a foundation for AI-driven visualisation techniques in XR. Participants' preferences for visualisation techniques varied significantly based on context and environmental factors.

Integrating qualitative analysis into human-AI interaction research is essential for designing systems that align with human needs and societal values. Our qualitative research uncovered behaviour patterns, preferences, and key factors influencing these behaviours, providing valuable insights for future design development and user experience optimisation.

The results highlight the importance of maintaining user autonomy, ensuring transparent AI systems to build trust, and considering context when selecting visualisation techniques. Future work should focus on implementing these techniques in working systems, refining UI designs, and exploring additional use cases to encourage broader research within the community.

\begin{acks}
\end{acks}

\bibliographystyle{ACM-Reference-Format}
\bibliography{references_pub}

\appendix

\end{document}